\documentclass[%
reprint,
]{revtex4-1}

\usepackage{graphicx}
\usepackage{dcolumn}
\usepackage{bm}
\usepackage{amsmath}

\begin{document}
\title{A measurable counterpart to the optical concept of an object}

\author{Minghui Zhang}
\email{zmh@ahu.edu.cn}
\affiliation{School of Physics and Material Science, Anhui University, No.111
Jiulong Road, Hefei, 230601, P. R. China}

\author{Jiawen Li}
\affiliation{Department of Precision Machinery and Precision Instrumentation, University of Science and Technology of China,  Hefei, 230026, P. R. China}

\date{\today}

\begin{abstract}
The traditional optical concept for the object does not provide an experimental feasibility to speak for itself, due to the fact that no measuring instrument catches up with the fluctuation of light fields. Using the theory of coherence, we give it a measurable counterpart, and thus  to give a satisfactory explanation of what the concept has said. To our knowledge it is the first time to valuate an object in optics in the terms of observable quantities.  As a useful example, the applicability to obtain the full knowledge of an object under the partial coherent illuminations is suggested. This work hit the thought which had been advanced by E. Wolf that one measures the correlation function rather than the optical fields themselves. The suggested applicability is a supplementary to the  recently found solution to the determination of phases of the diffracted beam.
\end{abstract}


\maketitle

\section{\label{sec:intro}Introduction}
When one talks about an object in optics, he would refer to  the effects which it exerts on the illuminating light fields. To valuate these effects  one associates the object to a complex amplitude $A(\boldsymbol\rho,\omega)e^{j\theta(\boldsymbol\rho,\omega)}$ with a term like \emph{optical complex transmittance} (OCT) or \emph{optical complex reflectance} (OCR). In standard textbooks (e.g. in Ref. \cite{goodman_fo}) it has traditionally been defined as when the object was incident by a monochromatic electromagnetic wave with a frequency of $\omega$, it would modulate the transmitted/refleced  wave front at the position of $P$ specified by a vector $\boldsymbol\rho$ in the way of multiplying $A(\boldsymbol\rho, \omega)$ times to its amplitude while applying the phase delay of $\theta(\boldsymbol\rho, \omega)$ to its argument.

However when one investigates the problem in terms of observable quantities \cite{wolf_observable_1954}, the concept just mentioned is questionable. On one hand, the time-independent sources of monochromatic wave in the optical region of the electromagnetic spectrum are never encountered in real world, and all optical fields, for both their amplitudes and their phases have always been undergoing random fluctuations \cite{goodman_so}. Such field fluctuations, comparing to the respond speed of the instrument, are too rapid to be measured, so one can never compare the  value of the optical fields just before and just after it have transmitted through/refleted by the objects. On the other hand, what the quantity of being directly measured  is the intensity, which is an averaged square modulus of the light fields over a time span approximately as long as the reciprocal of an instrument's cut-off frequency. During this time span the phase knowledge  of the fluctuating optical fields must have been washed out due to the time integral performed by the measuring instrument, so one is never able to obtain the phase knowledge of the fields before and after it have transmitted through/refleted by the objects.  For these reasons the present defined concept has not \emph{automatically} provided an applicable environment to  speak for itself.

In this letter, we justify the original optical concept of object by providing a measurable counterpart to it. Meanwhile, as a useful example, we suggests that when the illuminating sources are \emph{partially} coherent, the acquisition to the knowledge of $A(\boldsymbol\rho,\omega)e^{j\theta(\boldsymbol\rho,\omega)}$ can  be achieved. It is a supplementary to the  recently found solution to the determination of phases of the diffracted beam \cite{wolf_prl_2009}.

The paper is organized as follows. In Sec. \ref{sec:joint_intensity}, we go over the joint intensity and show it is a measurable complex quantity. Then, in Sec. \ref{sec:cross_spec} we theoretically approve that the cross-spectral density function is the joint intensity when an optical fields was being filtered at the frequency of $\omega$. After that, the effects that an object exerts on the cross-spectral density will be presented in Sec. \ref{sec:modulation}. It is the physical significance this paper gives. Finally in Sec.\ref{sec:Conclusions} we give a measurable counterpart to justify the original optical concept of the \emph{object} by combing the three conclusions drawn in Sec.\ref{sec:joint_intensity} - Sec.\ref{sec:modulation}. In addition, an applicability is also suggest in the last section.

\section{\label{sec:joint_intensity}Joint intensity is a measurable quantity}
For simplicity, we only discuss the object in terms of OCT, although the discussions for OCR  are straightforward.
\begin{figure}\includegraphics
[bb=47 206 583 393,scale=0.45]
{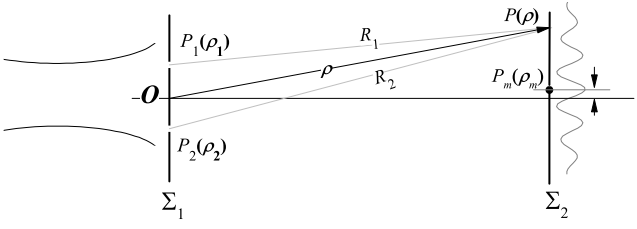}\caption{\label{fig:1}A schemed setup based on Young¡¯s interference experiment. The position of point $P$ is specified by a vector $\boldsymbol\rho$. In the setup an optical beam is incident on an opaque planar screen
$\Sigma_{1}$ with $P_{1}$ and $P_{2}$ opened. The diffraction pattern is observed in screen $\Sigma_{2}$. The contrast degree of the pattern is $\boldsymbol\gamma$, and the position of the pattern's maxima is at point $P_{m}$ specified by $\boldsymbol\rho_{\boldsymbol{m}}$. }
\end{figure}
The notation relating to the following discussion is given by Fig.\ref{fig:1}. We begin by going over the mutual coherence function \cite{born_n_wolf} defined as
\begin{align}
\Gamma ( {{{\boldsymbol{\rho }}_{\boldsymbol{1}}},{{\boldsymbol{\rho }}_{\boldsymbol{2}}},\tau } ) \equiv \langle {V^{\ast}( {{{\boldsymbol{\rho }}_{\boldsymbol{1}}},t} )V( {{{\boldsymbol{\rho }}_{\boldsymbol{2}}},t + \tau } )} \rangle . \label{eq:mutual_coherence_function}
\end{align}
In which $V(\boldsymbol{\rho}, t)$ is known as the complex analytic function \cite{mandel_wolf_rev_mod_phys} (also in Ref. \cite{wolf_newbook} page 92) representing a fluctuating field at a point $P$ specified by a vector $\boldsymbol{\rho}$ at time $t$, the asterisk  is a complex conjugate operator, and the angular brackets denote the ensemble average. Its  case for $\tau = 0$,
\begin{align}
J( {{{\boldsymbol{\rho }}_{\boldsymbol{1}}},{{\boldsymbol{\rho }}_{\boldsymbol{2}}}} ) \equiv \Gamma ( {{{\boldsymbol{\rho }}_{\boldsymbol{1}}},{{\boldsymbol{\rho }}_{\boldsymbol{2}}},0} ), \label{eq:joint_intensity}
\end{align}
is being called joint intensity between $P_{1}$ and $P_{2}$. When $\boldsymbol{\rho}_{\boldsymbol{1}}=\boldsymbol{\rho}_{\boldsymbol{2}}=\boldsymbol{\rho}$, the expression
\begin{align}
I( {\boldsymbol{\rho }} ) \equiv J( {{\boldsymbol{\rho }},{\boldsymbol{\rho }}} ),\label{eq:intensity}
\end{align}
defines a quantity which was directly measurable, formally named as the intensity at a position of $P$.  After a lengthy but straightforward calculation, the diffraction patterns $I(\boldsymbol\rho)$ in in screen $\Sigma_{2}$, can be carried out as \cite{wolf_1955}:
\begin{align}
\begin{gathered}
  I\left( \boldsymbol{\rho }  \right) = \left[ {{I^{\left( 1 \right)}}\left( {\boldsymbol{\rho }} \right) + {I^{\left( 2 \right)}}\left( {\boldsymbol{\rho }} \right)} \right] \hfill \\
  \begin{array}{*{20}{c}}
  {}&{}&{}
\end{array} \times \left\{ {1 + \frac{{2\sqrt {{I^{\left( 1 \right)}}\left( {\boldsymbol{\rho }} \right){I^{\left( 2 \right)}}\left( {\boldsymbol{\rho }} \right)} }}{{\left[ {{I^{\left( 1 \right)}}\left( {\boldsymbol{\rho }} \right) + {I^{\left( 2 \right)}}\left( {\boldsymbol{\rho }} \right)} \right]\sqrt {I\left( {{{\boldsymbol{\rho }}_{\boldsymbol{1}}}} \right)I\left( {{{\boldsymbol{\rho }}_{\boldsymbol{2}}}} \right)} }}} \right. \hfill \\
  \begin{array}{*{20}{c}}
  {}&{}&{}
\end{array}\left. { \times \left| {J\left( {{{\boldsymbol{\rho }}_{\boldsymbol{1}}},{{\boldsymbol{\rho }}_{\boldsymbol{2}}}} \right)} \right|\cos \left[ {\kappa  + \delta \left( {\boldsymbol{\rho }} \right)} \right]} \right\} \hfill \\
\end{gathered} \label{eq:intensity_distibution}
\end{align}
In which
\begin{align}\kappa=\arg{J\left( {{\boldsymbol{\rho} _{\boldsymbol{1}}},{\boldsymbol{\rho} _{\boldsymbol{2}}}} \right)}\end{align}\label{eq:kapa}is the argument of $J\left( {{\boldsymbol{\rho} _{\boldsymbol{1}}},{\boldsymbol{\rho} _{\boldsymbol{2}}}} \right)$,
$\delta\left( \boldsymbol{\rho}  \right)  = \frac{{2\pi }}{{\bar \lambda }}\left(| {{{\boldsymbol{\rho }}} - {{\boldsymbol{\rho }}_{\boldsymbol{2}}}} | - | {{{\boldsymbol{\rho }}} - {{\boldsymbol{\rho }}_{\boldsymbol{1}}}} |\right)$
is the phase difference at $P$ resulted from the difference of optical distances form $P$ to $P_{2}$ and $P$ to $P_{1}$. The latter can be simplified as
\begin{align}\delta\left( \boldsymbol{\rho}  \right)=\frac{2{\pi}D}{\bar{\lambda}d}\boldsymbol{\rho}\label{eq:phase_delay_simp}\end{align}
under the paraxial approximation, and where $\bar\lambda$ is the central wavelength for the light filed under the narrow band assumption. In Eq. \ref{eq:intensity_distibution},  $I^{(k)}(\boldsymbol{\rho }) (k = 1,2)$, are intensity distributions on $\Sigma_{2}$ when only $P_{k} (k = 1,2)$ was opened.  On the scale of $\frac{\bar{\lambda}d}{D}$, $I^{(k)}(\boldsymbol{\rho }) (k = 1,2)$  varied so slowly comparing to cosine  that they can be seen as constance in the terms of
\begin{align}{I^{\left( k \right)}}\left( {\boldsymbol{\rho }} \right) = {I^{\left( k \right)}},\left( {k = 1,2} \right).\label{eq:I_k}\end{align}
On substituting Eqs. \ref{eq:phase_delay_simp}, \ref{eq:I_k} From  Eq. \ref{eq:intensity_distibution}, one sees the observed pattern in in in screen $\Sigma_{2}$ has a sinusoidal profile in the form of
\begin{align}
\begin{gathered}
  I\left( {\boldsymbol{\rho }} \right) = \left[ {{I^{\left( 1 \right)}} + {I^{\left( 2 \right)}}} \right]\left\{ {1 + \frac{{2\sqrt {{I^{\left( 1 \right)}}{I^{\left( 2 \right)}}} }}{{\left[ {{I^{\left( 1 \right)}} + {I^{\left( 2 \right)}}} \right]\sqrt {I\left( {{{\boldsymbol{\rho }}_{\boldsymbol{1}}}} \right)I\left( {{{\boldsymbol{\rho }}_{\boldsymbol{2}}}} \right)} }}} \right. \hfill \\
  \begin{array}{*{20}{c}}
  {}&{}&{}
\end{array}\left. { \times \left| {J\left( {{{\boldsymbol{\rho }}_{\boldsymbol{1}}},{{\boldsymbol{\rho }}_{\boldsymbol{2}}}} \right)} \right|\cos \left[ {\kappa  + \frac{{2\pi D}}{{\overline \lambda  d}}{\boldsymbol{\rho }}} \right]} \right\}. \hfill \\
\end{gathered}
\label{eq:intensity_simp}\end{align}
Eq. \ref{eq:intensity_simp} shows that the contrast $\gamma$ of the observed pattern is:
\begin{align}
\gamma = \frac{{2\sqrt {{I^{\left( 1 \right)}}{I^{\left( 2 \right)}}} }}{{\left[ {{I^{\left( 1 \right)}} + {I^{\left( 2 \right)}}} \right]\sqrt {I\left( {{\boldsymbol{\rho} _{\boldsymbol{1}}}} \right)I\left( {{\boldsymbol{\rho} _{\boldsymbol{2}}}} \right)} }}\left| {J\left( {{\boldsymbol{\rho} _{\boldsymbol{1}}},{\boldsymbol{\rho} _{\boldsymbol{2}}}} \right)} \right|,
\label{eq:contrast}\end{align}
as defined by
\begin{align}
\gamma  \equiv \frac{{{I_{\max }}( {\boldsymbol{\rho }} ) - {I_{\min }}( {\boldsymbol{\rho }} )}}{{{I_{\max }}( {\boldsymbol{\rho }} ) + {I_{\min }}( {\boldsymbol{\rho }} )}}.\label{eq:contrast_df}
\end{align}
Eq. \ref{eq:contrast_df} also provides an experimental way to obtain the contrast of the diffracted patterns. This fact combined with Eq. \ref{eq:contrast}  tells one that the modules $|J\left( {{\boldsymbol{\rho} _{\boldsymbol{1}}},{\boldsymbol{\rho} _{\boldsymbol{2}}}} \right)|$ is obtainable because $\gamma ,{I^{\left( k \right)}},I\left( {{{\boldsymbol{\rho }}_k}} \right),\left( {k = 1,2} \right)$ in Eq. \ref{eq:contrast} can all be determined by the experiment. Besides, from Eq. \ref{eq:intensity_simp} one can see that the argument of $J\left( {{\boldsymbol{\rho} _{\boldsymbol{1}}},{\boldsymbol{\rho} _{\boldsymbol{2}}}} \right)$, namely $\kappa$, can also be determined by observing the pattern's maxima position $P_{\boldsymbol{m}}$ in a way of
\begin{align}\kappa  =  - \frac{{2\pi D}}{{\overline \lambda  d}}{{\boldsymbol{\rho }}_m},\label{eq:kapa_measure}\end{align}
where ${{\boldsymbol{\rho }}_{\boldsymbol{m}}}$ stands for the position of the  pattern's maxima position $P_{\boldsymbol{m}}$ .In a word, from Eqs. \ref{eq:contrast} - \ref{eq:kapa_measure} one can say that \emph{the joint intensity (Eq.\ref{eq:joint_intensity}) is a measurable quantity }since all terms in the right side of equation,
\begin{align}J\left( {{{\boldsymbol{\rho }}_{\boldsymbol{1}}},{{\boldsymbol{\rho }}_{\boldsymbol{2}}}} \right) = \gamma \frac{{\left[ {{I^{\left( 1 \right)}} + {I^{\left( 2 \right)}}} \right]\sqrt {I\left( {{{\boldsymbol{\rho }}_{\boldsymbol{1}}}} \right)I\left( {{{\boldsymbol{\rho }}_{\boldsymbol{2}}}} \right)} }}{{2\sqrt {{I^{\left( 1 \right)}}{I^{\left( 2 \right)}}} }}e^{j\kappa},\label{eq:joint_int_cacu}\end{align}
are obtainable. This is the first conclusion for latter reference. Although a special case when $I^{(1)}=I^{(2)}$ and $I\left( {{{\boldsymbol{\rho }}_{\boldsymbol{1}}}} \right) = I\left( {{{\boldsymbol{\rho }}_{\boldsymbol{2}}}} \right)$ had discussed to the measurability for the mutual coherence function in Ref. \cite{born_n_wolf}, a more general case in form of Eq. \ref{eq:joint_int_cacu} is still needed for the later discussions.

\section{\label{sec:cross_spec}Cross-spectral density function and its relation to the joint intensity}
Suppose the field is statistically stationary, at least in the wide sense, the Fourier transform of Eq. \ref{eq:mutual_coherence_function}
\begin{align}
W( {{{\boldsymbol{\rho }}_{\boldsymbol{1}}},{{\boldsymbol{\rho }}_{\boldsymbol{2}}},\omega } ) \equiv \frac{1}{{\sqrt {2\pi } }}\int_{ - \infty }^\infty  {\Gamma ( {{{\boldsymbol{\rho }}_{\boldsymbol{1}}},{{\boldsymbol{\rho }}_{\boldsymbol{2}}},\tau } ){e^{j\omega \tau }}d\tau } \label{eq:cross_spectual_intensity}
\end{align}
is known as the cross-spectral density function \cite{mandel_wolf_rev_mod_phys,wolf_newbook}. It can be expressed \cite{wolf_oc_1981,wolf_josa_1982,Tervo_josa} in the coherent-mood representation \cite{mandel_wolf_1981}(also see Ref. \cite{wolf_newbook}, page 214, and Ref. \cite{Herrero}.):
\begin{align}
W( {{{\boldsymbol{\rho }}_{\boldsymbol{1}}},{{\boldsymbol{\rho }}_{\boldsymbol{2}}},\omega } ) = \sum\limits_n {{\lambda _n}( \omega  )\varphi _n^*( {{{\boldsymbol{\rho }}_{\boldsymbol{1}}},\omega } ){\varphi _n}( {{{\boldsymbol{\rho }}_{\boldsymbol{2}}},\omega } )} ;\label{eq:cross_spec_intensity_expan}
\end{align}
in which $\lambda_{n}(\omega)>0$ may be shown to be the eigenvalues and and $\varphi(\boldsymbol{\rho },\omega)$ the eigenfunctions of the homogeneous Fredholm integral equation:
\begin{align}
\int_D {W( {{{\boldsymbol{\rho }}_{\boldsymbol{1}}},{{\boldsymbol{\rho }}_{\boldsymbol{2}}},\omega } ){\varphi _n}( {{{\boldsymbol{\rho }}_{\boldsymbol{1}}},\omega } )} {{\boldsymbol{d}}^{\boldsymbol{3}}}{{\boldsymbol{\rho}}_{\boldsymbol{1}}} = {\lambda _n}( \omega  ){\varphi _n}( {{{\boldsymbol{\rho }}_{\boldsymbol{2}}},\omega } ),\label{eq:homogeneous_integral}
\end{align}
Where $D$ is the domain containing all pairs of $P_{1}$ and $P_{2}$. Its eigenfunctions may be taken to form an orthonormal set as
\begin{align}
\int_D {\varphi _m^*( {{\boldsymbol{\rho }},\omega } ){\varphi _n}( {{\boldsymbol{\rho }},\omega } )} {{\boldsymbol{d}}^{\boldsymbol{3}}}{\boldsymbol{\rho }} = {\delta _{mn}},\label{eq:orthonormal}
\end{align}
$\delta_{mn}$ being the Kronnecker symbol ($\delta_{mn} =1$ when $m = n$, $\delta_{mn} = 0$ when $m \neq n$.).

Let us consider a situation for a beam which has passed through a narrow filter with central frequency of $\omega$, then the fluctuating fields at the point of $P$ may be seen as a random process among an ensemble of $\{U( {{\boldsymbol{\rho }},\omega } )\}$ \cite{wolf_newbook}, in which $U( {{\boldsymbol{\rho }},\omega } )$ is a frequency dependent sample function. Let $U( {{\boldsymbol{\rho }},\omega } )$  to be a monochromatic realization \cite{explian} of the ensemble, and the fields at the point of $P_{1}$ and $P_{2}$ can be constructed as
\begin{align}
U\left( {{{\boldsymbol{\rho}}_{\boldsymbol{1}}},\omega } \right) = \sum\limits_m {{a_m}\left( \omega  \right){\varphi _m}\left( {{{\boldsymbol{\rho}}_{\boldsymbol{1}}},\omega } \right)}, \label{eq:U_expan_a}
\end{align}
and
\begin{align}
U\left( {{{\boldsymbol{\rho}}_{\boldsymbol{2}}},\omega } \right) = \sum\limits_n {{a_n}\left( \omega  \right){\varphi _n}\left( {{{\boldsymbol{\rho}}_{\boldsymbol{2}}},\omega } \right)}; \label{eq:U_expan_b}
\end{align}
due to the orthonormal properties of $\varphi_{m,n}(\boldsymbol{\rho },\omega)$. The $a_{m,n}(\omega)$ are random coefficients satisfied
\begin{align}
\langle {a_m^*( \omega  ){a_n}( \omega  )} \rangle  = {\alpha _n}{\delta _{mn}}.\label{eq:random_coef}
\end{align}
Now considering the joint intensity:
\begin{align}
{J^{( \omega  )}}( {{{\boldsymbol{\rho }}_{\boldsymbol{1}}},{{\boldsymbol{\rho }}_{\boldsymbol{2}}}} ) \equiv \langle {{U^*}( {{{\boldsymbol{\rho}}_{\boldsymbol{1}}},\omega } )U( {{{\boldsymbol{\rho}}_{\boldsymbol{2}}},\omega } )} \rangle \label{eq:jointintensity_afrer_filter}
\end{align}
(superscript notes the beam has passed through a narrow filter with central frequency of $\omega$ ).
By referring to Eqs. \ref{eq:mutual_coherence_function} and \ref{eq:joint_intensity} and Eqs. \ref{eq:U_expan_a} and \ref{eq:U_expan_b}, one has
\begin{align}
\begin{array}{l}
{J^{( \omega  )}}( {{{\boldsymbol{\rho }}_{\boldsymbol{1}}},{{\boldsymbol{\rho }}_{\boldsymbol{2}}}} ) = \langle {[ {\sum\limits_m {a_m^*( \omega  ){\varphi _m}( {{{\boldsymbol{\rho }}_1},\omega } )} } ]}\\
\begin{array}{*{20}{c}}
{}&{}
\end{array} { \times [ {\sum\limits_n {{a_n}( \omega  ){\varphi _n}( {{{\boldsymbol{\rho }}_{\boldsymbol{2}}},\omega } )} } ]} \rangle \\
\begin{array}{*{20}{c}}
{}&{}
\end{array} = \sum\limits_m {\sum\limits_n {\langle {a_m^*( \omega  ){a_n}( \omega  )} \rangle {\varphi _m}( {{{\boldsymbol{\rho }}_1},\omega } ){\varphi _n}( {{{\boldsymbol{\rho }}_{\boldsymbol{2}}},\omega } )} }
.\end{array}
\label{eq:jointintensity_expan}
\end{align}
On substituting Eq. \ref{eq:random_coef} from Eq.\ref{eq:jointintensity_expan} and by letting
\begin{align}
{\alpha _n} = \sqrt {{\lambda _n}} ,\label{eq:squair_root_lmd}
\end{align}
one can derive joint intensity of Eq. \ref{eq:jointintensity_expan} out to be
\begin{align}{J^{( \omega  )}}( {{{\boldsymbol{\rho }}_{\boldsymbol{1}}},{{\boldsymbol{\rho }}_{\boldsymbol{2}}}} ) = \sum\limits_n {{\lambda _n}{\varphi _n}( {{{\boldsymbol{\rho }}_{\boldsymbol{1}}},\omega } ){\varphi _n}( {{{\boldsymbol{\rho }}_{\boldsymbol{2}}},\omega } )} .\label{eq:jointintensity_expan_2}
\end{align}
Comparing Eq. \ref{eq:jointintensity_expan_2} with Eq. \ref{eq:cross_spec_intensity_expan} one may get
\begin{align}{J^{( \omega  )}}( {{{\boldsymbol{\rho }}_{\boldsymbol{1}}},{{\boldsymbol{\rho }}_{\boldsymbol{2}}}} ) =  W( {{{\boldsymbol{\rho }}_{\boldsymbol{1}}},{{\boldsymbol{\rho }}_{\boldsymbol{2}}} },\omega ).\label{eq:joint_intensity0}
\end{align}
To this end we are able to give a physical meaning to Eq. \ref{eq:cross_spectual_intensity}. That is \emph{the cross-spectral density function $W( {{{\boldsymbol{\rho }}_{\boldsymbol{1}}},{{\boldsymbol{\rho }}_{\boldsymbol{2}}},\omega } )$ is the joint intensity (Eq. \ref{eq:joint_intensity}) when an optical fields was being filtered at the central frequency of $\omega$. } This is the second conclusion of this paper \cite{alternative}. Similar to Eq. \ref{eq:intensity}, the equal position case of Eq. \ref{eq:joint_intensity0},
\begin{align}I(\boldsymbol\rho,\omega) = J^{(\omega)}( {{\boldsymbol{\rho }},{\boldsymbol{\rho }} } )\label{eq:joint_intensity1}
\end{align}
defines the intensity at point $P$ after the light was thus filtered.

\section{\label{sec:modulation}The modulation to the cross-spectral density function by an object}
In free space, the mutual coherence function (Eq.\ref{eq:mutual_coherence_function}) satisfies the wave equations \cite{wolf_selected} (also known as Wolf equation \cite{saleh_prl_2005}),
\begin{align}
\nabla _j^2\Gamma  - \frac{1}{{{c^2}}}\frac{{{\partial ^2}\Gamma }}{{\partial {\tau ^2}}} = 0,\begin{array}{*{20}{c}}
{}&{}
\end{array}j = 1,2;\label{eq:wolf_eq}
\end{align}
where $\nabla^{2}_{j}$ is the Laplacian operator with respect to $\boldsymbol{\rho }_{\boldsymbol{j}}$. By substitute the inverse form of Eq. \ref{eq:cross_spectual_intensity},
\begin{align}
\Gamma ( {{{\boldsymbol{\rho }}_{\boldsymbol{1}}},{{\boldsymbol{\rho }}_{\boldsymbol{2}}},\tau } ) = \frac{1}{{\sqrt {2\pi } }}\int_{ - \infty }^\infty  {W( {{{\boldsymbol{\rho }}_{\boldsymbol{1}}},{{\boldsymbol{\rho }}_{\boldsymbol{2}}},\omega } ){e^{ - j\omega \tau }}d\omega } ,\label{eq:invers_fourier}
\end{align}
into Eq. \ref{eq:wolf_eq}, one obtains
\begin{align}
\nabla _j^2W( {{{\boldsymbol{\rho }}_{\boldsymbol{1}}},{{\boldsymbol{\rho }}_{\boldsymbol{2}}},\omega } ) + {k^2}W( {{{\boldsymbol{\rho }}_{\boldsymbol{1}}},{{\boldsymbol{\rho }}_{\boldsymbol{2}}},\omega } ) = 0.\label{eq:Helmholtz_W}
\end{align}
On substituting Eq.\ref{eq:cross_spec_intensity_expan} from Eq.\ref{eq:Helmholtz_W} with respect to $j = 2$, multiplying the resulted equation by $\varphi_{m}(\boldsymbol{\rho }_{\boldsymbol{2}},\omega)$, integrating its both sides with respect to the variable ${{{\boldsymbol{\rho }}_{\boldsymbol{2}}}}$  over the domain $D$,
and using the orthonormality relation Eq.\ref{eq:orthonormal}, one obtains
\begin{align}
{\nabla ^2}{\varphi _n}( {{{\boldsymbol{\rho }}_{\boldsymbol{1}}},\omega } ) + {k^2}{\varphi _n}( {{{\boldsymbol{\rho }}_{\boldsymbol{1}}},\omega } ) = 0.\label{eq:Helmholtz_varphi}
\end{align}
Now we can see that time-independent function $\varphi _n( {{{\boldsymbol{\rho }}_{\boldsymbol{1}}},\omega } )$  is a monochromatic electromagnetic wave which had been traditionally used to define the concept of complex objects  because it obeys Helmholtz equation (Eq. \ref{eq:Helmholtz_varphi}). Now, attaching an object at $P_{1}$ and applying the traditional concept, Eq.\ref{eq:U_expan_a} changes into
\begin{align}
U'( {{{\boldsymbol{\rho }}_{\boldsymbol{1}}},\omega } ) = \sum\limits_m {{a_m}( \omega  )[ {A( {{{\boldsymbol{\rho }}_{\boldsymbol{1}}},\omega } ){e^{j\theta ( {{{\boldsymbol{\rho }}_{\boldsymbol{1}}},\omega } )}} \cdot {\varphi _m}( {{{\boldsymbol{\rho }}_{\boldsymbol{1}}},\omega } )} ]} .\label{eq:U_expan_2}
\end{align}
On substituting Eq.\ref{eq:U_expan_2} from Eq.\ref{eq:jointintensity_afrer_filter} for $U( {{{\boldsymbol{\rho }}_{\boldsymbol{1}}},\omega } )$ and by going over the deriving procedure from Eq.\ref{eq:U_expan_a} to Eq.\ref{eq:joint_intensity0}, one can get the cross-spectral density function for this situation as
\begin{align}
\begin{gathered}
  W'\left( {{{\boldsymbol{\rho }}_{\boldsymbol{1}}},{{\boldsymbol{\rho }}_{\boldsymbol{2}}},\omega } \right) = \left\langle {{{U'}_1}^*\left( {{{\boldsymbol{\rho }}_{\boldsymbol{1}}}} \right){U_2}\left( {{{\boldsymbol{\rho }}_2}} \right)} \right\rangle  \hfill \\
  \begin{array}{*{20}{c}}
  {}&{}&{}&{}& =
\end{array}A\left( {{{\boldsymbol{\rho }}_{\boldsymbol{1}}},\omega } \right){e^{j\theta \left( {{{\boldsymbol{\rho }}_{\boldsymbol{1}}},\omega } \right)}}W\left( {{{\boldsymbol{\rho }}_{\boldsymbol{1}}},{{\boldsymbol{\rho }}_{\boldsymbol{2}}},\omega } \right). \hfill \\
\end{gathered} \label{eq:W_prime}
\end{align}
Eq. \ref{eq:W_prime} states that \emph{When placing a complex object immediate after point $P_{1}$, it changes the cross-spectral density function of $W( {{{\boldsymbol{\rho }}_{\boldsymbol{1}}},{{\boldsymbol{\rho }}_{\boldsymbol{2}}},\omega } )$ into $W'( {{{\boldsymbol{\rho }}_{\boldsymbol{1}}},{{\boldsymbol{\rho }}_{\boldsymbol{2}}},\omega } )$.} This is the third conclusion  and it is also the physical significance of this paper. Here we note again that Eq. \ref{eq:U_expan_2} is the bridge between the gap of the traditional concept and a measurable one.

\section{\label{sec:Conclusions}A measurable counterpart to the optical concept of the object }
Combining the three conclusions drawn in Sec.\ref{sec:joint_intensity} - \ref{sec:modulation}, we can give a counterpart to the optical concept of an object from the perspective of measurable quantities: \emph{One can associate an object to a complex amplitude $A(\boldsymbol\rho,\omega)e^{j\theta(\boldsymbol\rho,\omega)}$, it means when an object is placing at the position of $P_{1}$ in the Young¡¯s interference experiment setup shown by Fig.\ref{fig:1}, it modules the measurable cross-spectral density function $W( {{{\boldsymbol{\rho }}_{\boldsymbol{1}}},{{\boldsymbol{\rho }}_{\boldsymbol{2}}},\omega } )$ in the way of multiplying $A(\boldsymbol\rho_{\boldsymbol{1}}, \omega)$ times to its amplitude while applying the phase delay of $\theta(\boldsymbol\rho_{\boldsymbol{1}}, \omega)$ to its argument.}

Meanwhile, through the above discussion, the applicability  to obtain the complex transmittance is suggested by the experimental setup of Fig. \ref{fig:1}: Filtering the incident partial coherent light by a narrowband filter with central frequency of $\omega$, then according to the first two conclusion of this paper, the cross-spectral density function of $W( {{{\boldsymbol{\rho }}_{\boldsymbol{1}}},{{\boldsymbol{\rho }}_{\boldsymbol{2}}},\omega } )$ and $W'( {{{\boldsymbol{\rho }}_{\boldsymbol{1}}},{{\boldsymbol{\rho }}_{\boldsymbol{2}}},\omega } )$, which equals to the joint intensity of Eq. \ref{eq:joint_intensity} before and after the complex object was attached to the point $P_{1}$, can be measured according to Eq. \ref{eq:joint_int_cacu} in a way of:
\begin{align}
\begin{gathered}
  W\left( {{{\boldsymbol{\rho }}_{\boldsymbol{1}}},{{\boldsymbol{\rho }}_{\boldsymbol{2}}},\omega } \right) = \left[ {{I^{\left( 1 \right)}}\left( \omega  \right) + {I^{\left( 2 \right)}}\left( \omega  \right)} \right] \hfill \\
  \begin{array}{*{20}{c}}
  {}&{}&{}&{}
\end{array} \times \frac{{\sqrt {I\left( {{{\boldsymbol{\rho }}_{\boldsymbol{1}}},\omega } \right)I\left( {{{\boldsymbol{\rho }}_2},\omega } \right)} }}{{2\sqrt {{I^{\left( 1 \right)}}\left( \omega  \right){I^{\left( 2 \right)}}\left( \omega  \right)} }}\gamma {e^{j\kappa \left( \omega  \right)}} ,\hfill \\
\end{gathered}
\end{align}
and
\begin{align}
\begin{gathered}
  W'\left( {{{\boldsymbol{\rho }}_{\boldsymbol{1}}},{{\boldsymbol{\rho }}_{\boldsymbol{2}}},\omega } \right) = \left[ {{{I'}^{\left( 1 \right)}}\left( \omega  \right) + {I^{\left( 2 \right)}}\left( \omega  \right)} \right] \hfill \\
  \begin{array}{*{20}{c}}
  {}&{}&{}&{}
\end{array} \times \frac{{\sqrt {I'\left( {{{\boldsymbol{\rho }}_{\boldsymbol{1}}},\omega } \right)I\left( {{{\boldsymbol{\rho }}_2},\omega } \right)} }}{{2\sqrt {{{I'}^{\left( 1 \right)}}\left( \omega  \right){I^{\left( 2 \right)}}\left( \omega  \right)} }}\gamma '{e^{j\kappa '\left( \omega  \right)}} .\hfill \\
\end{gathered}
\end{align}
In which the superscripts have the same meaning which we have already introduced when describing Eq. \ref{eq:intensity_simp}, and the prime ``$'$" denotes the values after the object being placed. By such information, one can obtain $A(\boldsymbol\rho_{\boldsymbol{1}}, \omega)^{ej\theta(\boldsymbol\rho_{\boldsymbol{1}}, \omega)}$ from
\begin{align}
A( {{{\boldsymbol{\rho }}_{\boldsymbol{1}}},\omega } ) = \left| {\frac{{W'( {{{\boldsymbol{\rho }}_{\boldsymbol{1}}},{{\boldsymbol{\rho }}_{\boldsymbol{2}}},\omega } )}}{{W( {{{\boldsymbol{\rho }}_{\boldsymbol{1}}},{{\boldsymbol{\rho }}_{\boldsymbol{2}}},\omega } )}}} \right|,
\end{align}
and
\begin{align}
\theta ( {{{\boldsymbol{\rho }}_{\boldsymbol{1}}},\omega } ) = \kappa '\left( \omega  \right) - \kappa \left( \omega  \right) .
\end{align}

In summary, we provide a measurable counterpart to the traditional imaginarily defined optical concept for the \emph{object}. Meanwhile the applicability to retrieval the complex knowledge of the object is suggested. This research hit the thought advanced by Wolf that one measures the correlation function rather than the optical fields themselves \cite{wolf_oc_2011, wolf_observable_1954,wolf_prl_2009}. The suggested applicability can be seen as a supplementary to the  recently found solution to the determination of phases of the diffracted beam \cite{wolf_prl_2009}.

\begin{acknowledgments}
One of the author (M.Z.) is sincerely grateful to Professor Emil Wolf for his helpful advice. The work was supported by National Science Foundation of China (No.51275502, 51205375), and China Postdoctoral Foundation funded project (No.2012M511416).
\end{acknowledgments}


\begin{thebibliography}{99}
\bibitem{goodman_fo}J. W. Goodman, Introduction to Fourier optics, 2 ed. (McGraw-Hill Companies, Colorado, 2005).
\bibitem{wolf_observable_1954}E. Wolf, ``Optics in terms of observable quantities," Nuovo Cimento \textbf{12}, 884-888 (1954).
\bibitem{wolf_prl_2009} E. Wolf, ``Solution of the Phase Problem in the Theory of Structure Determination of Crystals from X-Ray Diffraction Experiments," Phys. Rev. Lett. \textbf{103}, 075501-075503 (2009).
\bibitem{goodman_so}J. W. Goodman, Statistical Optics (John Wiley $\&$  Sons, Incorporated, Hoboken, 2000).
\bibitem{born_n_wolf}M. Born and E. Wolf, \emph{Principles of Optics}, Seventh(expanded) ed. (Cambridge University Press, New York, 1999).
\bibitem{mandel_wolf_rev_mod_phys}L. Mandel and E. Wolf, "Coherence Properties of Optical Fields," Rev. Mod. Phys. \textbf{37}, 231-287 (1965).
\bibitem{wolf_1955}E. Wolf, ``A Macroscopic Theory of Interference and Diffraction of Light from Finite Sources. II. Fields with a Spectral Range of Arbitrary Width," in Proc. R. Soc. Lond. A,  (1955), p. 246.
\bibitem{wolf_newbook}L. Mandel and E. Wolf, \emph{Optical Coherence and Quantum Optics} (Press Syndicate of the University of Cambridge, Cambridge, 1995).
\bibitem{explian}$U\left( {{\boldsymbol{r}},\omega } \right)$ is not a a Fourier component of the fluctuating field but is the space-dependent part of the statistical ensemble $\left\{ {V\left( {\boldsymbol{r},t} \right) = U\left( {{\boldsymbol{r}},\omega } \right){e^{ - j\omega t}}} \right\}$ of monochromatic realizations, all of frequency $\omega$. That is why $U\left( {{\boldsymbol{r}},\omega } \right)$ can be bounded. To appreciate such realizations, one can refer to Ref. \cite{wolf_2007}. Also, in Ref. \cite{Lahiri_2011}, authors refered to $U\left( {{\boldsymbol{r}},\omega } \right)$ as the \emph{associated field}.
\bibitem{wolf_2007}E. Wolf, \emph{Introduction to the Theory of Coherece and Polariztion of Light} (Cambridge University Press, New York, 2007).
\bibitem{Lahiri_2011} M. Lahiri and E. Wolf, ``Implications of complete coherence in the space-frequency domain," Opt. Lett. 36, 2423-2425 (2011).
\bibitem{alternative}An alternative derivation to this conclussion can be inferred form \cite{alternative_b}, although the authors haven't state it directly.
\bibitem{alternative_b}M. Lahiri and E. Wolf, ``Does a light beam of very narrow bandwidth always behave as a monochromatic beam?," Physics Letters A 374, 997-1000 (2010).
\bibitem{wolf_oc_1981}E. Wolf, ``New spectral representation of random sources and of the partially coherent fields that they generate," Opt. Commu. \textbf{38}, 3-6 (1981).
\bibitem{wolf_josa_1982}E. Wolf, ``New theory of partial coherence in the space-frequency domain. Part I: spectra and cross spectra of steady-state sources," J. Opt. Soc. Am. \textbf{72}, 343-351 (1982).
\bibitem{Tervo_josa}J. Tervo, T. Set\"{a}l\"{a}, and A. T. Friberg, ``Theory of partially coherent electromagnetic fields in the space-frequency domain," J. Opt. Soc. Am. A \textbf{21}, 2205-2215 (2004).
\bibitem{mandel_wolf_1981}L. Mandel and E. Wolf, ``Complete coherence in the space-frequency domain," Opt. Commu. \textbf{36}, 247-249 (1981).
\bibitem{Herrero}R. Mart¨ªnez-Herrero and P. M. Mej¨ªas, "Expansion of Coherence Functions of Stationary, Partially Coherent, Polychromatic Fields," Optica Acta: International Journal of Optics \textbf{29}, 187-195 (1982).
\bibitem{wolf_selected}E. Wolf, in \emph{Selected works of Emil Wolf with commentary} (World Scientific Publishing Co. Pte. Ltd, Singapore, 2001).
\bibitem{saleh_prl_2005}B. E. A. Saleh, M. C. Teich, and A. V. Sergienko, ``Wolf Equations for Two-Photon Light," Phys. Rev. Lett. \textbf{94}, 223601-223604 (2005).
\bibitem{wolf_oc_2011}E. Wolf, ``What kind of phases does one measure in usual interference experiments?," Opt. Commu. \textbf{284}, 4235-4236 (2011).

\end{thebibliography}
\end{document}